\documentclass[10pt, conference]{IEEEtran}

\IEEEoverridecommandlockouts
\usepackage{cite}
\usepackage{amsmath,amssymb,amsfonts}
\usepackage{algorithmic}
\usepackage{graphicx}
\usepackage{textcomp}
\usepackage{xcolor}
\usepackage[numbers,sort&compress]{natbib}
\usepackage[font=small,skip=4pt]{caption}
\usepackage{url}

\usepackage{cleveref}

\usepackage{booktabs} % For formal tables

\renewcommand{\IEEEbibitemsep}{0pt plus 2pt}
\makeatletter
\IEEEtriggercmd{\reset@font\normalfont\footnotesize}
\makeatother
\IEEEtriggeratref{1}

\usepackage[boxed,ruled]{algorithm2e}
\usepackage{multirow}
\usepackage{makecell}
\usepackage{wrapfig}
\usepackage{setspace}

\usepackage[left=1.6cm,right=1.6cm,top=1.6cm,bottom=1.9cm]{geometry}

\renewcommand\baselinestretch{0.883}
\IEEEoverridecommandlockouts
\IEEEpubid{\makebox[\columnwidth]{978-1-7281-2954-9/19/\$31.00 $\copyright$2019 IEEE \hfill} \hspace{\columnsep}\makebox[\columnwidth]{ }}

\begin{document}

%\title{\LARGE An Ultra-Efficient Memristor-Based DNN Framework with Structured Weight Pruning and Quantization Using ADMM \vspace{-2em}}

\title{\LARGE An Ultra-Efficient Memristor-Based DNN Framework with Structured Weight Pruning and Quantization Using ADMM \vspace{-0.5em}}

\author{\IEEEauthorblockN{Geng Yuan\text{$^\dagger$}}
\IEEEauthorblockA{\fontsize{8}{10.8}\textit{Northeastern University} \\
Boston, USA \\
yuan.geng@husky.neu.edu}
\and
\IEEEauthorblockN{Xiaolong Ma\text{$^\dagger$}}
\IEEEauthorblockA{\fontsize{8}{10.8}\textit{Northeastern University} \\
Boston, USA \\
ma.xiaol@husky.neu.edu}
\and
\IEEEauthorblockN{Caiwen Ding}
\IEEEauthorblockA{\fontsize{8}{10.8}\textit{Northeastern University} \\
Boston, USA \\
ding.ca@husky.neu.edu}
\and
\IEEEauthorblockN{Sheng Lin}
\IEEEauthorblockA{\fontsize{8}{10.8}\textit{Northeastern University} \\
Boston, USA \\
lin.sheng@husky.neu.edu}
\and
\IEEEauthorblockN{Tianyun Zhang}
\IEEEauthorblockA{\fontsize{8}{10.8}\textit{Syracuse University} \\
Boston, USA \\
tzhan120@syr.edu}
\and
\IEEEauthorblockN{Zeinab S. Jalali}
\IEEEauthorblockA{\fontsize{8}{10.8}\textit{Syracuse University} \\
Syracuse, USA \\
zsaghati@syr.edu}
\and
\IEEEauthorblockN{Yilong Zhao}
\IEEEauthorblockA{\fontsize{8}{10.8}\textit{Shanghai JiaoTong University} \\
Shanghai, China \\
ylzhao2018@gmail.com}
\and
\IEEEauthorblockN{Li Jiang}
\IEEEauthorblockA{\fontsize{8}{10.8}\textit{Shanghai JiaoTong University} \\
Shanghai, China \\
jiangli@cs.sjtu.edu.cn}
\and
\IEEEauthorblockN{Sucheta Soundarajan}
\IEEEauthorblockA{\fontsize{8}{10.8}\textit{Syracuse University} \\
Syracuse, USA \\
susounda@syr.edu}
\and
\IEEEauthorblockN{Yanzhi Wang}
\IEEEauthorblockA{\fontsize{8}{10.8}\textit{Northeastern University} \\
Boston, USA \\
yanz.wang@northeastern.edu}
}

\maketitle
\def\footnoterule{\relax%
  \kern-5pt
  \hbox to \columnwidth{\hfill\vrule width 1\columnwidth height 0.4pt\hfill}
  \kern4.6pt}
\makeatother
\newcommand\blfootnote[1]{%
  \begingroup
  \renewcommand\thefootnote{}\footnote{#1}%
  \addtocounter{footnote}{-1}%
  \endgroup
}

\blfootnote{\hspace{-3.5mm}$^\dagger$These authors contributed equally.}
\begin{abstract}
The high computation and memory storage of large deep neural networks (DNNs) models pose intensive challenges to the conventional Von-Neumann architecture, incurring substantial data movements in the memory hierarchy.
The memristor crossbar array has emerged as a promising solution to mitigate the challenges and enable low-power acceleration of DNNs. Memristor-based weight pruning and weight quantization have been seperately investigated and proven effectiveness in reducing area and power consumption compared to the original DNN model. However,  there has  been no systematic investigation of memristor-based neuromorphic computing (NC) systems considering both weight  pruning and weight quantization.  
In this paper, we  propose  an  unified and  systematic  memristor-based  framework  considering both structured weight pruning and weight quantization by incorporating {\em alternating direction method of multipliers} (ADMM) into DNNs training. We consider hardware constraints such  as  crossbar  blocks  pruning,  conductance  range,  and mismatch  between  weight  value  and  real  devices,  to  achieve high  accuracy  and  low  power  and  small  area  footprint. Our framework is mainly integrated by three steps, i.e., memristor-based  ADMM  regularized  optimization, masked mapping and retraining. Experimental results show that our proposed framework achieves 29.81$\times$ (20.88$\times$) weight compression ratio, with 98.38\% (96.96\%) and 98.29\% (97.47\%) power and area reduction on VGG-16 (ResNet-18) network where only have 0.5\% (0.76\%) accuracy loss, compared to the original DNN models. We share our models at anonymous link 
% \normalfont{\textcolor{blue}{\texttt{\url{http://bit.ly/2Jp5LHJ}}}} .
\textcolor{blue}{\url{http://bit.ly/2Jp5LHJ}}.

% Moreover, since several hardware constraints are incorporated into our framework that makes it hardware-friendly, and the framework can mitigate the mapping mismatch between trained model and hardware implementation. 
\end{abstract}

\section{Introduction}

With the rise of artificial intelligence, Deep Neural Networks have been widely used thanks to their high accuracy, excellent scalability, and self-adaptiveness~\cite{goodfellow2016deep}. DNN models are becoming
deeper and larger, and are evolving fast to satisfy the diverse characteristics of broad applications.
% such as face detection~\cite{li2015convolutional}, speech recognition~\cite{amodei2016deep}, scene parsing~\cite{zhao2017pyramid}, and autonomous driving~\cite{chen2015deepdriving}. 
% For example, the representative DNN models AlexNet~\cite{krizhevsky2012imagenet}, VGGNet~\cite{simonyan2014very}, and ResNet-50~\cite{he2016deep} of ImageNet
% dataset~\cite{deng2009imagenet} consist of 60.9 million (M), 138 M, and 25.6 M parameters, respectively. 
The high
computation and memory storage of DNN models pose intensive challenges to the conventional Von-Neumann architecture, incurring substantial data movements in memory hierarchy.

To achieve high performance and energy efficiency, hardware acceleration of DNNs is intensively studied both in academia and industry~\cite{ding2017c,wang2018ultrahigh,Ding_2018,Ma_2018,shrestha2019approx,8489619,8697495,li2019admm}. 
% Most of these DNN accelerators focus on inference, which is a forward progress of input from the first layer to the last. 
DNN model compression techniques, including weight pruning~\cite{han2015learning,wen2016learning,zhang2018adam,ma2019resnet,ye2019progressive,niu201926ms} and weight quantization~\cite{park2017weighted,wu2016quantized,lin2019toward}, are developed to facilitate hardware acceleration by reducing storage/computation in DNN inference with negligible impact on accuracy. However, as Moore's law is coming to an end~\cite{waldrop2016chips}, the acceleration of the conventional Von-Neumann architecture is limited to some extent.
% Recently, researchers have proposed novel hardware architectures to accelerate deep neural networks (DNNs) algorithms~\cite{merolla2014million,chen2014dadiannao,jouppi2017datacenter} using standard CMOS devices since conventional processing units such as CPUs and GPUs do not support these algorithms well. 
% However, as Moore's law is coming to an end~\cite{waldrop2016chips}, new challenges have been brought on these hardware acceleration in DNNs. 

To further mitigate the intensive
computation and memory storage of DNN models, the next-generation device/circuit technologies beyond CMOS and novel computing architectures beyond the traditional Von-Neumann machine are investigated. The crossbar array of the recently discovered memristor devices (i.e., memristor crossbar) can be utilized to perform matrix-vector multiplication in the analog domain and solve systems of linear equations in $O(1)$ time complexity~\cite{chua1971memristor,Yuan2017}. 
% A pioneering work on \textbf{weight pruning} of DNNs was done by Han et al.~\cite{han2015learning}, which is an iterative heuristic method, directly removing the zeros in weights or remove those with small absolute values. It achieves a 9$\times$ reduction in the number of weights of AlexNet model (for ImageNet dataset).
Ankit et al.~\cite{ankit2017trannsformer} implemented weight pruning techniques to NC systems using memristor crossbar arrays, which reduces the area (energy) consumption
compared to the original network. However, for hardware implementations on on-chip neuromorphic computing systems, there are several limitations: (i) unbalanced workload; (ii) extra memory footprint on indices; (iii) irregural memory access. These will cause the circuit overheads in hardware implementations. To address these limitations, Wang. et al.~\cite{wang2017group} proposed group connection deletion, which prunes connections to reduce routing congestion between crossbar arrays.

On the other hand, Zhang. et al.~\cite{zhang2014quantized} discussed the effectiveness of using the quantized conductance in memristor in multi-level logics. 
Song. et al.~\cite{song2017quantization} investigated the generation of quantization loss in the memristor-based NC systems and its impacts on computation accuracy, and proposed a regularized offline learning method that can minimize the impact of quantization loss during neural network mapping.
Weight quantization can mitigate hardware  imperfection  of memristor including state  drift and process variations, caused  by the imperfect fabrication process or by the device feature itself.
% Xia et al.~\cite{xia2016switched} presented a technique to reduce the overhead of Digital-to-Analog Converters (DACs)/Analog-to-Digital Converters (ADCs) in resistive random-access memory (ReRAM) NC systems. They first normalized the data, and then quantized intermediary data to 1-bit value. This can be directly used as the analog input for ReRAM crossbar and, hence, avoids the need of DACs.  

Because weight pruning and weight quantization techniques leverage different sources
of redundancy, they may be combined to achieve higher DNN compression. However, there has been no
systematic investigation of this effect in the
memristor-based NC systems considering both weight pruning and weight quantization. In this paper, we propose an unified and systematic memristor-based framework considering both structured weight pruning and weight quantization, by incorporating ADMM into DNNs training. We consider hardware constraints such as crossbar blocks pruning,  conductance  range,  and  mismatch  between  weight  value and  real  devices, to achieve high accuracy and low power and small area footprint. Our proposed framework can better mitigate the inaccuracy caused by the hardware imperfection compared to only weight quantization method~\cite{zhang2014quantized, song2017quantization}. It contains \textit{memristor-based ADMM regularized  optimization}, \textit{masked mapping} and \textit{retraining steps}, which can guarantee the solution feasibility (satisfying all constraints) and provide high solution quality (maintaining test accuracy) at the same time. The contributions
of this paper include:

\begin{itemize}
    \item We systematically investigate the combination of structured weight pruning and weight quantization techniques leveraging different sources of redundancy, to achieve higher DNN compression ratio and low power and area in the domain of memristor-based NC systems. 
    \item We adopt ADMM, an effective optimization technique for general and non-convex optimization problems, to jointly optimize weight pruning and weight quantization problems during training for higher model accuracy.
    
\end{itemize}

We evaluate our proposed framework on different networks. Experimental  results  show  that our  proposed  framework  can  achieve  29.81$\times$(20.88$\times$)  weight compression ratio, with 98.38\% (96.96\%) and 98.29\% (97.47\%) power and area reduction on VGG-16 (ResNet-18) network where only have 0.5\% (0.76\%) accuracy loss, compared to the original DNN models.

\section{Background on Memristors}
\label{sec:2}

%------------------------------------------------------------
%------------------  Section 3.1  ---------------------------
% \subsection{Memristor}
\subsection{Memristor Crossbar Model}

\begin{figure} [!h]
     \centering
     \vspace{-1.2em}
     \includegraphics[width=0.85\columnwidth]{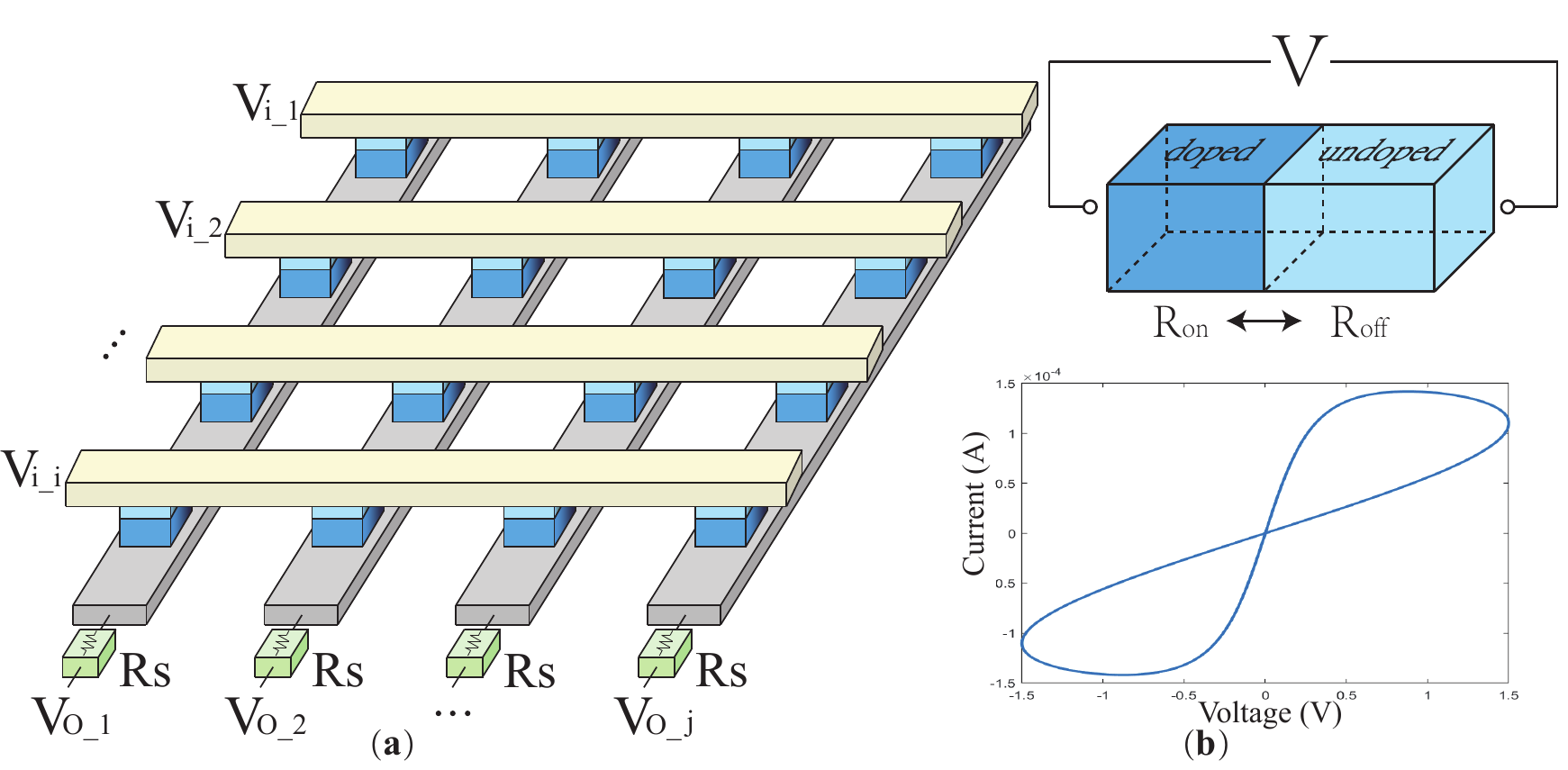}    
     \caption{(a) Memristor crossbar performs maxtrix-vector multiplication. (b) Memristor model and its $V$-$I$ curve.}
     \label{fig:mem}
 \end{figure}
 
Memristor has shown remarkable characteristics as one of the most promising emerging technologies as shown in Figure~\ref{fig:mem}~\cite{Radwan2010}. 
% It was implemented by HP Lab in 2008 and has been proved as the fourth fundamental element in circuit[\textcolor{blue}{ref}]. 
% The memristor consists of an electrically adjustable semiconductor thin-film structure. When an external voltage is applied to its two metal contacts, the memristor will works as a two terminal passive element.  By applying a positive (negative) voltage over a period of time, the resistance of the device, which is also known as the "memristance", will decrease (increase) as a result of the width of the doped region in the thin-film structure shrinks (extends). 
The memristor has many promising features, such as non-volatility, low-power, high integration density, and excellent scalability.
% , which makes it to be a desired candidate of the new memory device.
%Figure~\ref{fig:mem} shows a general structure of the memristor device. The memristor consists of an electrically adjustable semiconductor thin film of width D and two metal contacts. The thin file can be divided into a doped region that highly doped with oxygen vacancies and an undoped region~\cite{strukov2008missing}. When an external voltage is applied to its two metal contacts, the memristor will works as a two terminal passive element. And the doped region will act as the conductor with low resistance, while the undoped region has high resistance as the insulator. In Figure.1, the variable w represents the width of the doped region, then the resistance state R(w) of the memristor can be calculated by using Equation (\ref{mem-resistance}).
%\begin{equation}
%\label{mem-resistance}
%  R(w)=\big{(}R_{on}\cdot{w/D}+R_{off}\cdot{(1-w/D)}\big{)}
%\end{equation}
%By applying an external voltage bias across the memristor, the total resistance of the device will be changed. When a positive (negative) voltage is applied over a period of time, the resistance of the device will decrease (increase) as a result of the width of the doped region shrinks (extends). The variable $R_{on}$ and $R_{off}$ represents the minimum and maximum available resistance of the memristor, which can be reached when $w/D\rightarrow{0}$ and $w/D\rightarrow{1}$ respectively~\cite{memCharacteristics}. The programable and non-volatility feature makes memristor to be a desired candidate of the new memory device.
Memristors can be formed as a crossbar structure~\cite{memCrossbar1}, as shown in Figure \ref{fig:mem}. Each pair of horizontal Word-line (WL) and vertical Bit-line (BL) is connected across a memristor device. 
% When the whole WL is considered as an input vector, the entire crossbar structure can be treated as a matrix multiplication equation, where each memristor cell behaves as a multiplication factor. 
Given the input voltage vector $\textbf{v}_i$ and the weight matrix $\bf{W}$ which can be constructed by a preprogramed crossbar array, the matrix multiplication result $\textbf{v}_o$ can be easily obtained by measuring the current across the resistor $R_S$. By nature, the memristor crossbar array is attractive for matrix computations with high degree of parallelism which can achieve the time complexity of $O(1)$. Based on this superior feature, the memristor-based computing system can provide a promising solution to reduce the latency and improve the energy efficiency of neuromorphic computation.

\subsection{Hardware Imperfection of Memristor
Crossbars and Mitigation Techniques}

Hardware imperfection of Memristor is mainly caused by the imperfect fabrication process or by the device feature itself. These significant issues cannot be ignored in hardware design, which is different from the software-based system design.

\subsubsection{State Drift}
Memristor device consists of a thin-film structure, and the film is divided into two regions. One region is highly doped with oxygen vacancies and another region is an undoped. Applying an electric field across the device over time would lead to the migration of oxygen vacancies and change the memristance state, which is called state drift~\cite{Yang2008}. Thus, after a certain number of read operations, the resistance of the device will drift caused by the accumulative effect of applying the same direction voltage. As a result of the state drift effect, the imprecision will be incurred when the memristor's state drifts to the other state levels. 
% Researchers investigate to mitigate this effect through (i) performing periodic refreshing, (ii) retention time engineering for memristors, and (iii) adopting coarse-grained quantization/discretization on the resistance values of memristor device (sometimes even using binary (0/1) quantization). 

\subsubsection{Process Variation.} Process variation is also phenomenal as the process technology scales to nanometer level. It mainly comes from the line-edge roughness, oxide thickness fluctuations, and random dopant variations that affect the memristor device performance~\cite{ roughness}. The process variation will cause the hardware non-ideal behavior, which usually means the accuracy degradation~\cite{xiaProVar}.
% Researchers attempt to mitigate the impact of process variations through coarse-grained quantizations/discretization, adopting extra memristor cells, BL/WL engineering, etc..

It can be observed that quantization on resistance values plays an important role in dealing with hardware
imperfections. However, the prior work on mitigating the effect of hardware imperfections are mainly
{\em ad hoc}, lacking a systematic, algorithm-hardware co-optimization framework to improve overall resilience. 
% Considering these issues, besides our structure pruning framework, we also proposed a optimized framework using quantization method, which can 
Our proposed framework can mitigate the inaccuracy caused by the hardware imperfection, while achieves high hardware efficiency as well.

% \textcolor{red}{(add how to transfer to GEMM)}

% By applying the memristor crossbar structure and using the in-memory computing mechanism, the latency of the neuromorphic computation can be significantly reduced, and the power efficiency can be boosted at the same time.
%------------------  Section 3.3  ---------------------------
\section{A Unified and Systematic Memristor-Based Framework for DNNs}

The memristor crossbar structure has shown promising features in neuromorphic computing systems compared to the traditional CMOS technologies\cite{ankit2017trannsformer}. However, as DNN goes deeper and deeper, the massive weight computation and weight storage introduce severe challenges in neuromorphic computing system hardware implementations. On the other hand, to systematically address hardware imperfections of memristor crossbars, in this paper, we propose a integrated memristor-based framework 

\subsection{Unified and Systematic Memristor-Based Framework using ADMM}
\subsubsection{Connection to ADMM}
ADMM~\cite{boyd2011distributed} is a powerful optimization tool, by decomposing
an original problem into two subproblems that can be solved separately and iteratively.
% ADMM is conventionally utilized to accelerate the convergence of convex optimization problems and
% enable distributed optimization. 
% The optimality and fast convergence rate of ADMM have been proven for
% convex problems~\cite{ouyang2013stochastic}. 
% As a special property, ADMM can effectively deal with a subset of combinatorial constraints and yields optimal (or at least high quality) solutions~\cite{}. 
% Our key observation is that the associated constraints in the DNN weight pruning and quantization belong to this subset of combinatorial constraints, which indicates a great potential of ADMM in solving the optimization problems during DNN training.
Consider the
optimization problem
${\rm min}_{\bf{x}}~~ f(\bf{x})$ + $g(\bf{x})$. In ADMM, the problem is first re-written as

\begin{equation}
\label{admm}
  \underset{ \{{\bf{x,z}}\}}{\text{min}}~~f(\bf{x})+g(\bf{z}), ~~{\rm subject~ to~} \bf{x}=\bf{z}
\end{equation}

Next, by using augmented Lagrangian~\cite{boyd2011distributed}, the above problem is decomposed into two subproblems on $\bf{x}$ and $\bf{z}$. The first is ${\rm min}_{\bf{x}} ~~f(\bf{x})$ + $q_1(\bf{x})$, where $q_1(\bf{x})$ is a quadratic function. As $q_1(\bf{x})$ is convex, the complexity
of solving subproblem 1 is the same as minimizing $f(\bf{x})$. Subproblem 2 is ${\rm min}_{\bf{z}} ~~g(\bf{z})$ + $q_2(\bf{z})$, where $q_2(\bf{z})$ is again a quadratic function. The two subproblems will be solved iteratively
until convergence is achieved ~\cite{ouyang2013stochastic}.

% ADMM is conventionally utilized to accelerate the convergence of convex optimization problems and
% enable distributed optimization. The optimality and fast convergence rate of ADMM have been proven for
% convex problems~\cite{ouyang2013stochastic}. As a special property, ADMM can effectively deal with a subset of combinatorial constraints and yields optimal (or at least high quality) solutions~\cite{hong2016convergence}. Our key observation is that the associated constraints in the DNN weight pruning and quantization belong to this subset of combinatorial constraints, which indicates a great potential of ADMM in solving the optimization problems during DNN training.
\subsubsection{Unified Memristor-Based Framework}
There is a difficulty in using ADMM directly due to the non-convex nature of the objective function for DNN training, and thereby lacking of any guarantees on solution feasibility and solution quality. It becomes even more challenging when incorporating ADMM into training the memristor-based DNN model, where we need to consider hardware constraints such as crossbar blocks pruning, conductance range, and mismatch between weight value and real devices. To overcome this challenge, we proposed an unified memristor-based framework including \textit{memristor-based ADMM regularized  optimization}, \textit{masked mapping} and \textit{retraining steps}, which can guarantee the solution feasibility (satisfying all constraints) and provide high solution quality (maintaining test accuracy) at the same time.

%------------------------------------------------------------
%------------------  Section 4.3  ---------------------------
% To obtain the \textcolor{red}{memristor-based model} with structured pruned and quantized weights as we mentioned above, we used an integrated algorithm which included three main steps, the \textit{\textcolor{blue}{memristor-based ADMM regularized  optimization}}, \textit{masked mapping} and \textit{retraining}.

First, the {\textit{memristor-based ADMM regularized  optimization}} starts from a pre-trained DNN model without compression. Consider an $N$-layer DNNs, sets of weights and biases of the $i$-th (CONV or FC) layer are denoted by ${\bf{W}}_{i}$ and ${\bf{b}}_{i}$, respectively. And the \textit{loss function} of the \textit{N}-layer DNN is denoted by $f \big( \{{\bf{W}}_{i}\}_{i=1}^N, \{{\bf{b}}_{i} \}_{i=1}^N \big)$.
Combining the task of memristor-based structured pruning and weight quantization, the overall problem is defined by
\begin{equation}
\small
\label{original}
\begin{aligned}
& \underset{ \{{\bf{W}}_{i}\},\{{\bf{b}}_{i} \}}{\text{minimize}}
& & f \big( \{{\bf{W}}_{i}\}_{i=1}^N, \{{\bf{b}}_{i} \}_{i=1}^N \big),
\\ & \text{subject to}
& & {\bf{W}}_{i}\in {\bf{P}}_{i}, \; {\bf{W}}_{i}\in {\bf{Q}}_{i}, \; i = 1, \ldots, N.
\end{aligned}
\end{equation}
Given the value of $\alpha_{i}$, the set
% ${{\bf{P}}_{i}=\{{\bf{W}}_{i}|card(supp({\bf{W}}_{i}))\leq\alpha_{i}\}}$ 
${\bf{P}}_{i}=\{{\bf{W}}_{i}|$ the number of non-zero structured weights in $\bf{W}_{i}\leq\alpha_{i}\}$ 
reflects the constraint for memristor-based structured weight pruning.
% where 'card' refers to cardinality and 'supp' refers to the support set. 
Elements in ${\bf{P}}_{i}$ are the solution of ${\bf{W}}_{i}$ satisfying the number of non-zero elements (after structured pruning and memristor crossbar mapping) in ${\bf{W}}_{i}$ which is limited by $\alpha_{i}$ for layer $i$. Similarly, elements in ${\bf{Q}}_{i}$ are the solutions of ${\bf{W}}_{i}$, in which elements in ${\bf{W}}_{i}$ assume one of ${q_{i,1}, q_{i,2}, \cdots, q_{i,M_{i}}}$ (memristor state values), where $M_i$ denotes the number of available quantization level in layer $i$. Please note that the $q_{i,j}$ value indicates the $j$-th quantization level in layer $i$, and $q_{i,j}\in[-cond_{max},-cond_{min}] \cup [cond_{min},cond_{max}]$, where $cond_{min},$ $cond_{max}$ are the minimum and maximum valid conductance value of a specified memristor device. More specifically, we use indicator functions to incorporate the memristor-based structured pruning and weight quantization constraints into the objective function, which are

\begin{minipage}{0.48\linewidth}
\small
\begin{eqnarray*}g_{i}({\bf{W}}_{i})=
\begin{cases}
 0 & \text { if } {\bf{W}}_{i}\in {\bf{P}}_{i}, \\ 
 +\infty & \text { otherwise, }
\end{cases}
\end{eqnarray*}
\end{minipage}
\begin{minipage}{0.47\linewidth}
\small
\begin{eqnarray*}h_{i}({\bf{W}}_{i})=
\begin{cases}
 0 & \text { if } {\bf{W}}_{i}\in {\bf{Q}}_{i}, \\ 
 +\infty & \text { otherwise, }
\end{cases}
\end{eqnarray*}
\end{minipage}

for $i = 1, \ldots, N$. Then the original problem (\ref{original}) can be equivalently rewritten as 
\begin{equation}
\small
\label{admm_form}
 \underset{ \{{\bf{W}}_{i}\},\{{\bf{b}}_{i} \}}{\text{minimize}}
\ \ \ f \big( \{{\bf{W}}_{i} \}_{i=1}^N, \{{\bf{b}}_{i} \}_{i=1}^N \big)+\sum_{i=1}^{N} g_{i}({\bf{W}}_{i})+\sum_{i=1}^{N} h_{i}({\bf{W}}_{i}).
\end{equation}

We incorporate auxiliary variables ${\bf{Y}}_{i}$ and ${\bf{Z}}_{i}$, dual variables ${\bf{U}}_{i}$ and ${\bf{V}}_{i}$, then apply ADMM to decompose problem (\ref{admm_form}) into three subproblems. After that, We solve these subproblems iteratively until the convergence. Assume in iteration $k$, the first subproblem is
\vspace{-2.0em}

\begin{equation}
\small
\begin{aligned}
\label{equ7}
 \underset{ \{{\bf{W}}_{i}\},\{{\bf{b}}_{i} \}}{\text{minimize}}
\ \ \ & f \big( \{{\bf{W}}_{i} \}_{i=1}^N, \{{\bf{b}}_{i} \}_{i=1}^N \big)+\sum_{i=1}^{N} \frac{\rho_{i}}{2}  \| {\bf{W}}_{i}-{\bf{Y}}_{i}^{k}+{\bf{U}}_{i}^{k} \|_{F}^{2} \\& +  \sum_{i=1}^{N} \frac{\rho_{i}}{2}  \| {\bf{W}}_{i}-{\bf{Z}}_{i}^{k}+{\bf{V}}_{i}^{k} \|_{F}^{2}, \\
\end{aligned}
\end{equation}

The first term in problem (\ref{equ7}) is the differentiable (non-convex) loss function of the DNN, while the other quadratic terms are convex. As a result, this subproblem can be solved by stochastic gradient descent (e.g., the ADAM algorithm ~\cite{kingma2014adam}) similar to training the original DNN.

The solution $\{{\bf{W}}_{i}\}$ of subproblem 1 is denoted by $\{{\bf{W}}_{i}^{k+1}\}$. Then we aim to derive $\{{\bf{Z}}_{i}^{k+1}\}$ and $\{{\bf{Y}}_{i}^{k+1}\}$ in subproblem 2 and 3. Thanks to the characteristics in combinatorial constraints (the memristor-based structured pruning and weight quantization constraints), the optimal, analytical solution of the two subproblems are Euclidean projections. We can prove that the projections are: keeping $\alpha_i$ elements with largest magnitudes and setting remaining weights to zero; and to quantize every weight element to the closest valid memristor state value. Finally, we update dual variables ${\bf{U}}_{i}$ and ${\bf{V}}_{i}$ according to ADMM rule ~\cite{boyd2011distributed} and thereby complete the $k$-th iteration in \textit{memristor-based ADMM regularized  optimization}.

\textit{Masked Mapping and Retraining:} We first perform the Euclidean projection (mapping) on the derived ${\bf{W}}_{i}$ to guarantee that at most $\alpha_i$ values in each layer are non-zero. Since the zero weights will not be mapped on the memristor crossbar, we can mask the zero weights and retrain the DNN with non-zero weights using training sets. Particularly, this retraining step is similar to the \textit{ADMM regularized hardware optimization} step, but only the memristor weight quantization constraints need to be satisfied. In this way test accuracy can be partially restored.
% which can systematically reduce the number of weights and further optimized by quantizing the weight to fewer bits and satisfied hardware constraints, while achieve the state-of-the-art weight compression ratio without accuracy loss.
% We select structured pruning method instead of the irregular pruning. It is a unified framework that can create different type of sparsity such as filter-wise, channel-wise, and shape-wise sparsity, which can be efficiently mapped on the memristor-crossbar based hardware design.

% It has been demonstrated that ADMM can be a powerful tool for
% ADMM~\cite{boyd2011distributed} is an effective mathematical optimization for solving nonconvex optimization problems with combinatorial constraints. It decomposes the original problem into two subproblems
% that can be solved separately and efficiently~\cite{ouyang2013stochastic}. 

%%%%%%%%%%%%%%%%%%%%%%%%%%%%%%%%%%%%%%%%%%%%%%%%%%%%%%%%%%%%
%%%%%%%%%%%%%%%%%%%%  Section 4  %%%%%%%%%%%%%%%%%%%%%%%%%%%
\subsection{Memristor-Based Structured Pruning \& Quantization}

\subsubsection{Memristor-Based Structured Weight Pruning} 
\begin{figure} [!h]
     \centering
%      \vspace{-1.2em}
     \includegraphics[width=0.5\columnwidth]{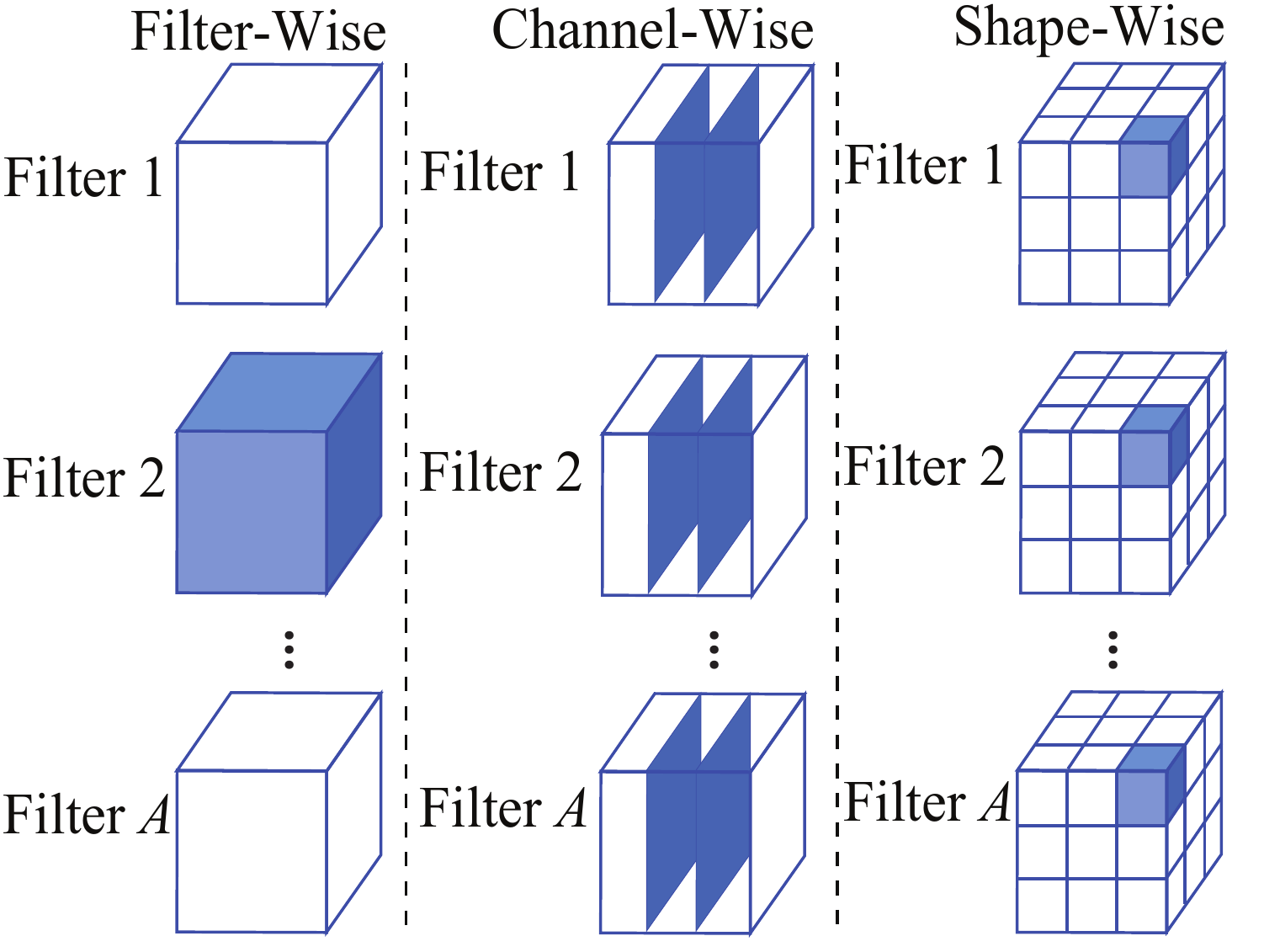}     \caption{Illustration of filter-wise,
channel-wise and shape-wise structured
sparsities.}
    \vspace{-0.2em}
     \label{fig:pruning}
\end{figure}

In order to be hardware-friendly, we use structured pruning method~\cite{wen2016learning} instead of the irregular pruning method~\cite{han2015learning} to reduce weights parameters. There are different types of structured sparsity, filter-wise sparsity, channel-wise sparsity, shape-wise sparsity as shown in Figure~\ref{fig:pruning}. 
In the proposed framework, We incorporate structured pruning in ADMM regularization, where memristor features are considered. Compared to~\cite{wang2017group}, our proposed method can better explore the sparsity on weight matrices, with negligible accuracy degradation, resulting in better area saving and lower power consumption. 

\begin{figure} [b]
     \centering
%      \vspace{-1.2em}
     \includegraphics[width=1\columnwidth]{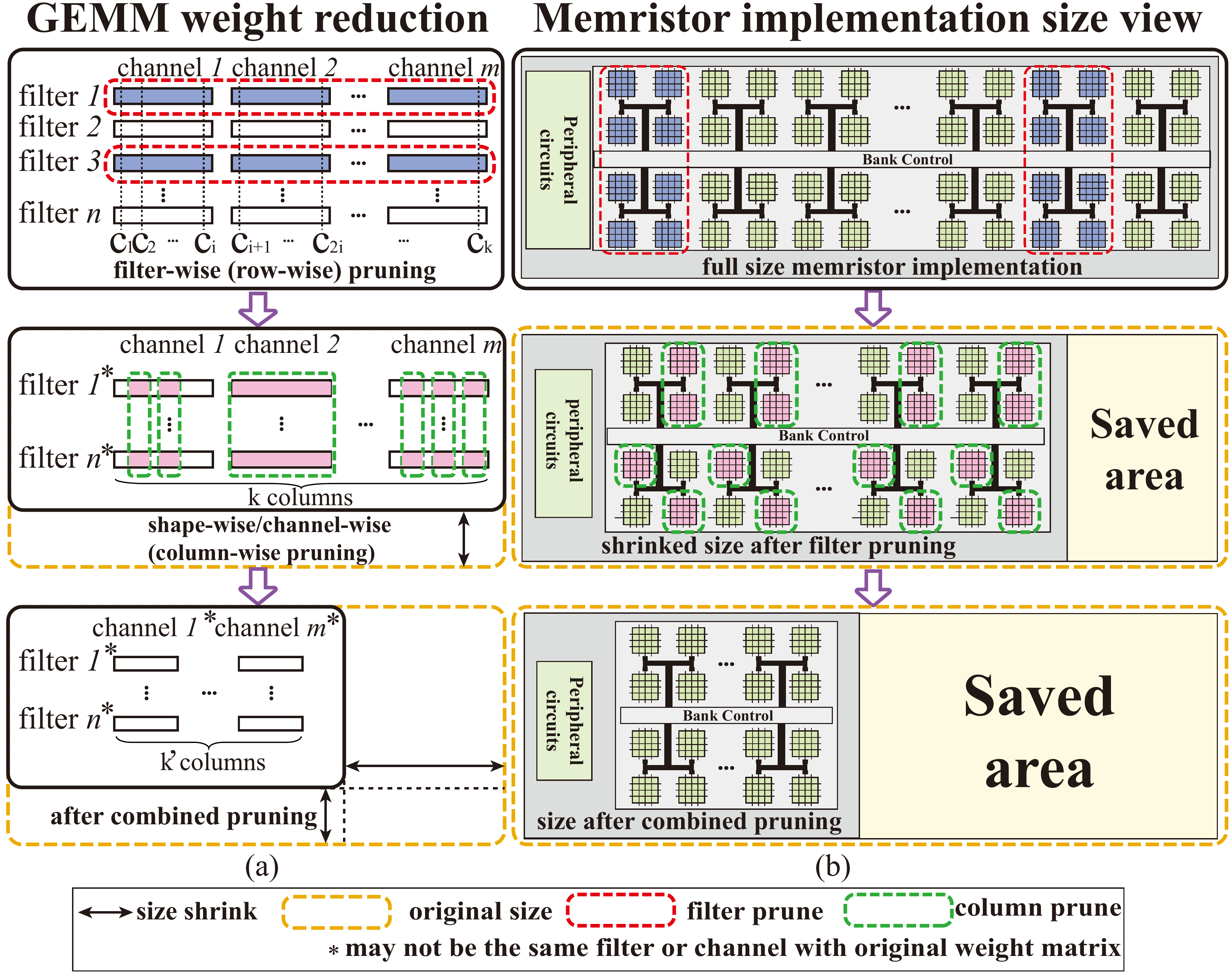}     
     \caption{Structured weight pruning and reduction of hardware resources}
     \label{fig:gemm_memristor_block}
\end{figure}

To better illustrate how structured pruning saves the memristor crossbars, we transform the weight tensors of a CONV layer to general matrix multiplication (GEMM) format~\cite{chetlur2014cudnn}. As shown in Figure \ref{fig:gemm_memristor_block} (a) (GEMM view), the
structured pruning corresponds to reducing rows or columns. The three structured sparsities, along with their combinations, will reduce the weight dimension in GEMM while maintaining a full matrix. Indices
are not needed and weight quantization will be better supported. 

Figure \ref{fig:gemm_memristor_block} (b) shows a memristor implementation size view and memristor crossbar area reduction on different types of sparsities. By applying filter (row) pruning and shape/channel (column) pruning, as shown in the top of Figure \ref{fig:gemm_memristor_block}(b), either blocks of memristor crossbar or numbers of memristor crossbars can be saved compared to the original design.

%The middle part shows some memristor crossbars within a block can be saved by the column-wise pruning, where the weights the same column represent the weights on the same position of all filters. The bottom part shows the total saved area after filter-wise (row-wise) pruning and column-wise pruning. Similarly, the channel-wise pruning can be achieved by column-wise pruning. By incorprating structured pruning, the weight matrix in each layer can be physically shrunken and the total cost of the memristor crossbars and the peripheral circuits can be significantly reduced. 

% \begin{wrapfigure}{O}{0.7\textwidth}
%     \begin{center}
%         \includegraphics[width=0.7\textwidth]{figs/gemm_memristor_block.pdf}
%     \end{center}
% \end{wrapfigure}

% , the filter-wise sparsity is equivalent to the row-sparsity on GEMM weight matrix. When a weight matrix with filter sparsity is mapped on to the memristor device, the memristor crossbars that suppose to hold the pruned filter weights can be removed. Similarly, the weights on a whole channel which refers to a group of weight columns on the weight matrix can be totally removed based on the channel sparsity. And the weights on a single column can be removed from the weight matrix regarding to the shape sparsity. 

\begin{figure} [t]
     \centering
%      \vspace{-1.2em}
     \includegraphics[width=0.7\columnwidth]{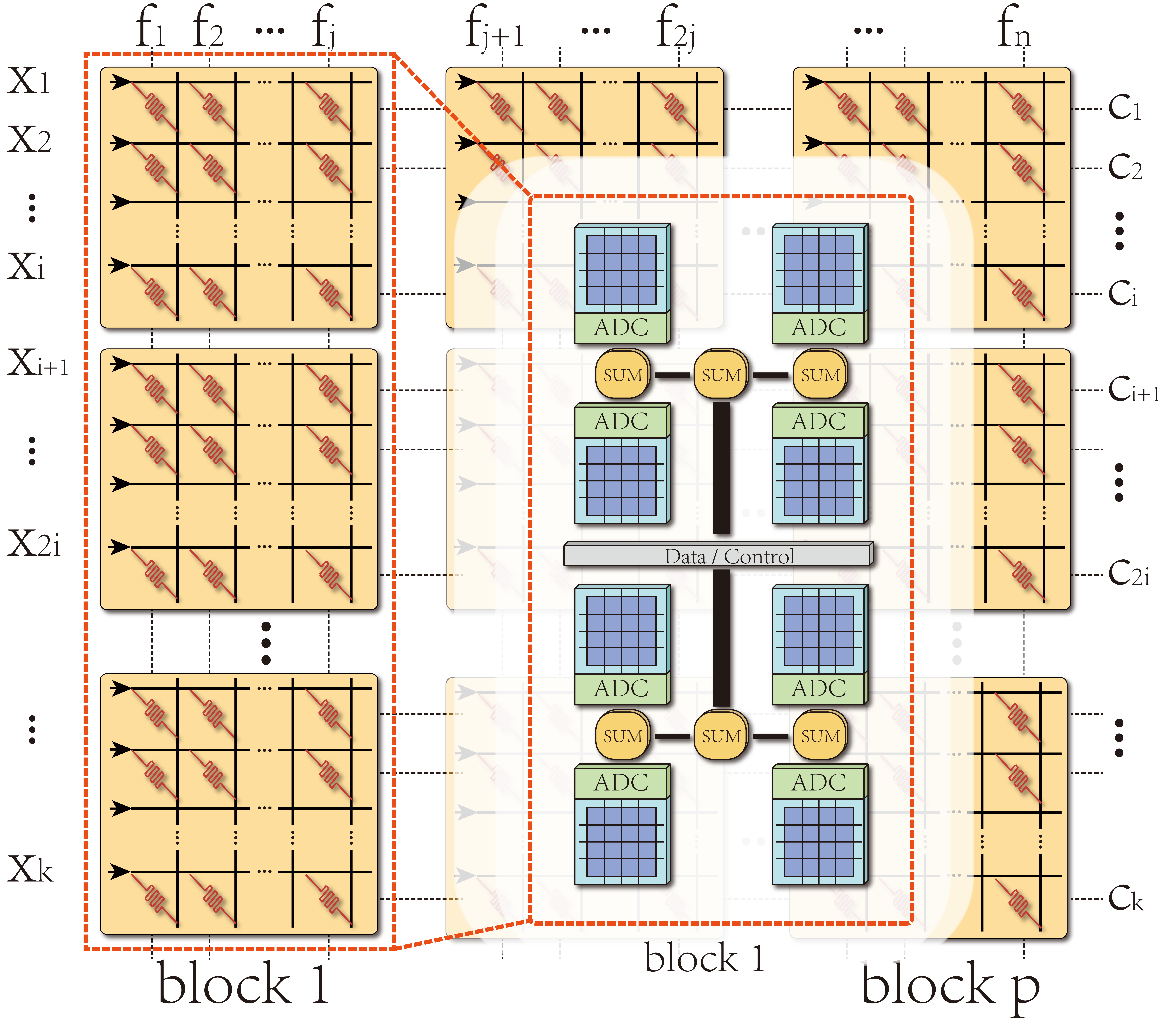}     \caption{Weights Mapping on Memristor crossbars}
     \label{fig:mapping_weights}
\end{figure}

The Figure~\ref{fig:mapping_weights} shows how we map the weight parameters on the memristor crossbars. As shown in the GEMM view in Figure \ref{fig:gemm_memristor_block} (a), assuming a CONV layer has $n$ filters, $m$ channels (including $k$ columns of weights), denoted as ${\bf{W}}\in\mathbb{R}^{n \times k}$. 
% Thus, the whole weight matrix has $n$ columns and $k$ rows, and there are $n\times k$ weights in total. 
Generally, the size of a single memristor crossbar is limited because the reading and writing error will increase by using larger crossbar size~\cite{Hu2016}. Thus, multiple memristor crossbars are used to accommodate the large size weight matrix. To maintain accuracy, the single memristor crossbar size in our design is no larger than 128$\times$64~\cite{Hu2018} and is identical for all DNN layers. As shown in Figure~\ref{fig:mapping_weights}, each crossbar has $i$ rows and $j$ columns. We use $X$ and $f$ to represent the inputs and filters, where $c$ represents the column of weights as shown in Figure \ref{fig:gemm_memristor_block} (a). Since there are $k$ weights in a filter, we need to use the columns at same position from at least $k/j$ different crossbars to store one filter's weights. Therefore, $j$ filters can be fully mapped on those crossbars as one block shown in Figure~\ref{fig:mapping_weights}. 
There are $n$ filters in total, therefore we need at least $p=n/j$ blocks to fully map the whole weight matrix (${\bf{W}}\in\mathbb{R}^{n \times k}$). Within each block, the outputs of each crossbar will be propagated through an ADC. Then We column-wisely sum the intermediate results of all crossbars.

%We use $X$ and $f$ to represent the inputs and filters, respectively. Each column indicates the weights on the same filter, and each row $C$ is the group of weights on the same position of different filters in Figure~\ref{fig:gemm_memristor_block} (a) (shape-wise). Assuming each memristor crossbar has $i$ rows and $j$ columns. Each column stores $i$ weights that from the same filter. Each row stores the $j$ weights on the same position from different $j$ filters. Generally, it is hard to use one column on one crossbar to store all the weights of a filter. Thus, a group of crossbars are used together to accommodate all the weights from the $j$ filters. Each group of crossbars is considered as a block and a H-tree structure is used for crossbar routing. \textcolor{red}{Within each block, the outputs on each crossbar will go through an ADC and then be column-wise added (SUM) with the results from the other crossbars. We need several blocks to store all filters' weights. In this way, the computation advantage of memristor crossbar could be fully utilized.}

%Note that the structured weight sparsity indicates that there are many zero weights structured located on the weight matrix. Since those zero weights do not make contributions to the computing results, they can be pruned away from the weight matrix. However, the pruned weights will be kept as zero during the training process, where they are not stored on the hardware when mapping the model to real hardware device. 

\subsubsection{Memristor-Based Weight Quantization}
% We take the hardware constraints into account when modeling the memristor-based framework. Please note 
The weights of the DNNs are represented by the conductance of memristors on memristor crossbars and the output of the memristor crossbars can be obtained by measuring the accumulated current. Due to the limited conductance range of the memristor devices, the weight values exceeding conductance range cannot be represented precisely. On the other hand, within the conductance range, accuracy loss also exists because of the mismatch between weight values and real memristor devices. 

%${\bf{Q}}_{i}= \{ \text{the value of every element}$ $\text{is one of the values in }~{q_1, q_2, \cdots, q_{M}} \}$

To mitigate the limitation of conductance range, we incorporate the conductance range constraint of the memristor device (i.e., ${\bf{Q}}_{i}\in[cond_{min},cond_{max}]$) into DNNs training. To mitigate the accuracy degradation caused by the weight mismatch, we incorporate the constraint of conductance state levels (i.e., ${q_{i,1}, q_{i,2}, \cdots, q_{i,M_{i}}}\in[cond_{min},cond_{max}]$) into DNNs training. Here set ${\bf{Q}}_{i}= \{$ the value of every element is one of the values in ${q_1, q_2, \cdots, q_{M}} \}$ to represent the constraint, and ${q_1, q_2, \cdots, q_{M}}$ are all available quantized states. Theoretically, the conductance of the memristor can be set to any states within its available range. In reality, the memristor conductance states are limited by the resolution that the peripheral write and read circuitry can provide. Generally speaking, more state levels require more sophisticated peripheral circuitry. In order to reduce the overheads caused by the peripheral circuitry and satisfy the robustness of the whole system, the conductance range will be quantized to several distinctive state levels and represented by discrete states. 

% Thus, we added a constraint into our hardware-oriented ADMM regularization \textcolor{red}{(ADMM regularized harware optimization)} process which can cluster weights to the quantized state levels of a specified memristor device during the training process. 

\begin{figure} [!h]
     \centering
%      \vspace{-1.2em}
     \includegraphics[width=0.6\columnwidth]{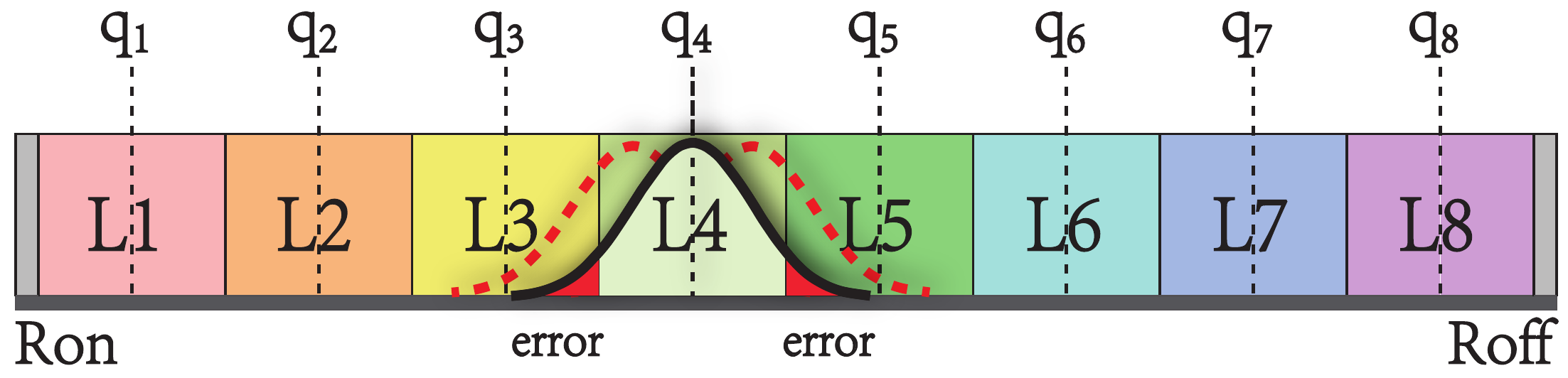}    
     \caption{Multi-level Memristor Storing 3-bit weight}
     \label{fig:multi-level}
 \end{figure}

Figure \ref{fig:multi-level} illustrates an example of an 8-level (3bits) memristor with linear conductance level (where it may behave as nonlinear in real designs~\cite{Lin2018}). The distribution curve shows a possible range that the memristor state might be actually set to, when the writing target state is $q_4$. An error will incur (hardware imprecision) when the actual written state is different from the target state level. In order to minimize the error caused by the hardware imprecision, in our constraint of conductance state levels, we set the quantized values as the mean value of each state level. To optimize the overall performance, the number of memristor state levels is also considered in our proposed framework. By quantizing the weights to fewer bits while maintain the overall accuracy, we can further improve the performance since fewer state levels provide longer distance for single state, resulting in better error resilience and reducing the hardware imprecision.

\begin{figure} [b]
     \centering
%      \vspace{-1.2em}
     \includegraphics[width=0.8 \columnwidth]{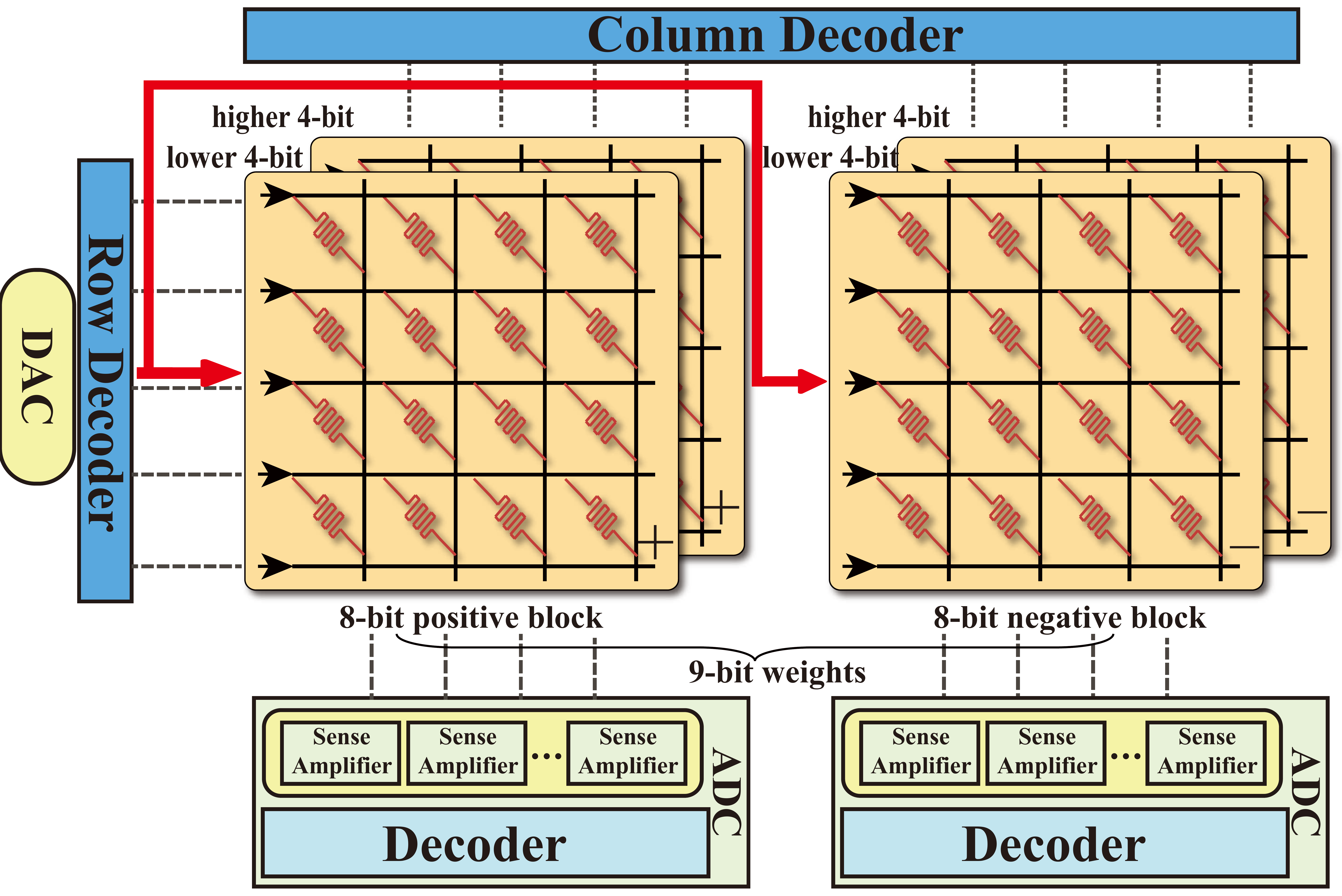}
     \caption{Represent Weights Using Multi-Memristor Crossbars}
     \label{fig:multi-memristor}
\end{figure}

Another advantage is the design area and power consumption can be reduced by quantizing the weights to fewer bits. According to the state-of-the-art design of neuromorphic computing using memristor, a practical assumption is that the memristor cell can represent 16 weight levels (4-bit weights) \cite{hardwareBits}. To ensure a relatively high accuracy, usually two (or more) memristors are bundled to represent weights with high resolution (more bits) \cite{PRIME}. On the other hand, since the memristor device only has positive conductance value while the weights are positive or negative, we use different memristor crossbars to represent the positive weights and negative weights separately. As an illustration in Figure~\ref{fig:multi-memristor}, a 9-bit weight value can be represented using a 8-bit positive block and a 8-bit negative block, where each of the 8-bit block is formed by two 4-bit memristor crossbars. In general, the cost of the ADCs and other peripheral circuits will grow exponentially for adding every extra bit precision. Thus, the overhead of the peripheral circuit can be significantly reduced by quantizing the weights to fewer bits. The total design area and power consumption can be reduced as well.

% Moreover, if the weights can be quantized to 5-bit or fewer, the half of the crossbars that are used to represent the higher bits and its corresponding peripheral circuit can be saved, thus }

\begin{table*}[t]
    \centering
    \caption{Structured Weight pruning results on multi-layer network on MNIST, CIFAR-10 datasets (\text{*}calculation is based on bolded results)}
    \renewcommand{\arraystretch}{1}
    \resizebox{0.9\textwidth}{!}{
        \begin{tabular}{c c c c c c c c} 
					\hline\hline
					\multicolumn{1}{|c|}{}& \multicolumn{4}{c}{\textbf{Structured Weight Pruning Statistics (9-bit)}} & \multicolumn{3}{|c|}{\textbf{Quantization \& Accuracy}} \\ 
					\hline
					\multicolumn{1}{|c|}{Method} & Original Accuracy & Pruned Accuracy & Crossbar Area Saved & Compression Ratio & \multicolumn{1}{|c}{7-bit} & 6-bit & \multicolumn{1}{c|}{5-bit} \\ 
					\hline
					\multicolumn{8}{c}{\textbf{MNIST}} \\
					\hline
					 Group Scissor~\cite{wang2017group} & 99.15\% & 99.14\% & 75.94\% & 4.16$\times$  & - & - & - \\ 
					\hline
					\multirow{3}{*}{\makecell{\textbf{our} \\ \textbf{LeNet-5}}} & \multirow{3}{*}{\textbf{99.17\%}} & \textbf{99.15\%} & \textbf{94.34\%} & \textbf{17.69$\times$}  & \textbf{98.97\%} & \textbf{99.03\%} & \textbf{99.03\%} \\ 
					 &  & 99.02\% & 97.30\% & 37.06$\times$    & 98.85\% & 98.82\% & 98.77\% \\ 
					 &  & 98.33\% & 99.05\% & 105.52$\times$    & 98.28\% & 98.10\% & 98.23\% \\ 
					\hline
					\multicolumn{8}{r}{\text{*}numbers of parameter reduced: \textbf{8.65K}}  \\
					\hline
					\multicolumn{8}{c}{\textbf{CIFAR-10}} \\
					\hline
					 Group Scissor~\cite{wang2017group} & 82.01\% & 82.09\% & 57.45\% & 2.35$\times$    & - & - & - \\ 
					\hline
					\multirow{3}{*}{\makecell{\textbf{our} \\ \textbf{ConvNet}}} & \multirow{3}{*}{\textbf{84.41\%}} & 84.55\% & 57.45\% & 2.35$\times$    & 84.18\% & 83.50\% & 80.81\% \\ 
					 &  & \textbf{84.53\%} & \textbf{65.87\%} & \textbf{2.93$\times$}   & \textbf{83.73\%} & \textbf{82.25\%} & \textbf{80.41\%} \\ 
					 &  & 83.58\% & 82.99\% & 5.88$\times$   & 83.00\% & 81.54\% & 78.18\% \\ 
					\hline
					\multirow{2}{*}{\makecell{\textbf{our} \\ \textbf{VGG-16}}} & \multirow{2}{*}{\textbf{93.70\%}} & 93.76\% & 89.26\% & 9.31$\times$ & 93.67\% & 93.64\% & 93.26\% \\
				% 	 &  & \textcolor{red}{93.27\%} & 95.87\% & 24.22$\times$ & 92.34\% & 91.68\% & \textcolor{red}{84.11\%} \\ 
					 &  & \textbf{93.36\%} & \textbf{96.65\%} & \textbf{29.81$\times$}    & \textbf{92.97\%} & \textbf{92.51\%} & \textbf{91.21\%} \\ 
					 \hline
					 \multirow{2}{*}{\makecell{\textbf{our} \\ \textbf{ResNet-18}}} & \multirow{2}{*}{\textbf{94.14\%}} & 93.79\% & 91.49\% & 11.75$\times$  & 93.68\% & 93.25\% & 92.92\% \\ 
				% 	 &  &  &  &     &  & 92.51\% & \textcolor{red}{85.87\%} \\ 
					 &  & \textbf{93.20\%} & \textbf{95.21\%} & \textbf{20.88$\times$}   & \textbf{93.13\%} & \textbf{92.65\%} & \textbf{92.44\%} \\ 
					 \hline
					 \multicolumn{8}{r}{\text{*}numbers of  parameter reduced on \textit{ConvNet}: \textbf{102.30K}, \textit{VGG-16}: \textbf{13.98M}, \textit{ResNet-18}: \textbf{10.46M}}  \\
				% 	 \hline
				% 	\multicolumn{8}{c}{\textbf{ImageNet ILSVRC-2012}} \\
				% 	\hline
				% 	\multirow{3}{*}{\makecell{\textbf{our's} \\ \textbf{AlexNet}}} & \multirow{3}{*}{\textbf{83.60\%}} & 83.58\% & 79.17\% & 4.81$\times$   & AA11 & AA12 & AA13 \\ 
				% 	 &  & 82.90\% & 86.36\% & 7.33$\times$   & AA21 & AA22 & AA23 \\ 
				% 	 &  & 81.76\% & 93.36\% & 15.05$\times$   & AA31 & AA32 & AA33 \\ 
				% 	 \hline
				% 	 \multicolumn{8}{r}{\text{*}total parameter reduced: \textbf{}}  \\
					 \hline\hline
					 
        \end{tabular}
    }
    \label{table:results}
\end{table*}

\begin{figure} [!h]
     \centering
%      \vspace{-1.2em}
     \includegraphics[width=0.8\columnwidth]{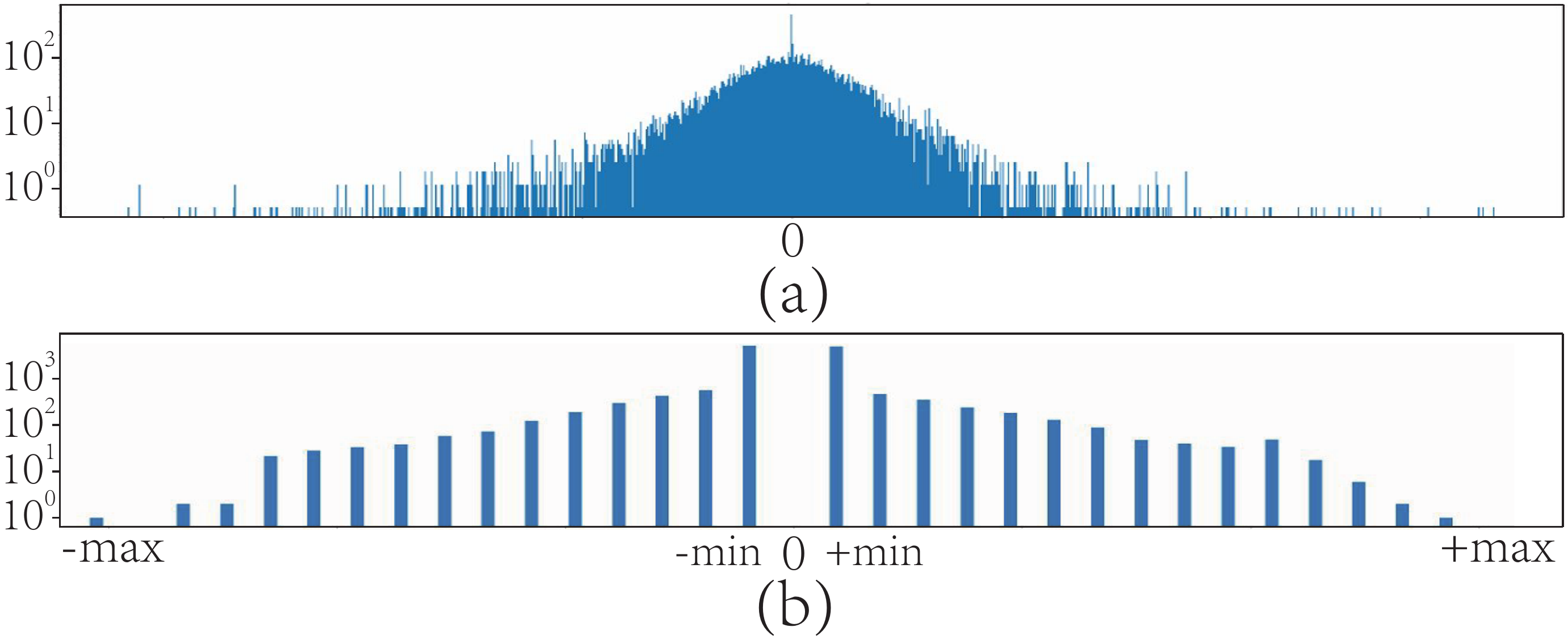}    
     \caption{Distribution of the weights: (a)before quantization, (b) after 5-bit quantization}
     \label{fig:ResNet_quant.pdf}
 \end{figure}

Figure~\ref{fig:ResNet_quant.pdf} shows the weights distribution of a CONV layer in ResNet18 using CIFAR-10 dataset before (b) and after (a) quantization, after structured weight pruning. For a 5-bit quantization using our proposed method, the weights are quantized into 32 different levels within memristor's valid conductance range. 
% The conductance range can be configured based on different type of memristors.

\section{experimental results}
In this section, we evaluate our systematic structured weight pruning and weight quantization framework on MNIST dataset using LeNet-5 network structure and CIFAR-10 dataset using ConvNet (4 CONV layers and 1 FC layer), VGG-16 and ResNet-18 network structures. All models are designed using PyTorch API and oriented to match the memristor's physical characteristics. Our hardware performance results such as power consumption, area cost of the memristor device and its peripheral circuits are simulated by using NVSim~\cite{Dong2012} and our MATLAB model. In our memristor model, $R_{min}=1M\Omega$, $R_{max}=10M\Omega$, with 4-bit precision, and the peripheral circuits are using 45nm technology. We use 128$\times$64 crossbar size on ResNet-18 and VGG-16, where ConvNet and LeNet-5 uses 32$\times$32 crossbar size.
The experiments are done on an eight NVIDIA GTX-2080Ti GPUs server. 

In this work, multiple 9-bit non-pruned models on different networks are used as our original DNN models, and results of structured weight pruning using our original DNN models show that, on memristor LeNet-5 model, we achieve 17.69$\times$ weight reduction without accuracy loss, 37.06$\times$ weight reduction with negligible accuracy loss and 105.52$\times$ weight reduction within 1\% accuracy loss. Meanwhile we shrink memristor crossbar area by more than 94\%. On muti-layers CNN for CIFAR-10, we achieve a higher accuracy compared with~\cite{wang2017group}. On deeper neural network structures such as VGG-16 and ResNet-18, we manage to compress each model unprecedentedly by 29.81$\times$ with negligible accuracy loss and 20.88$\times$ within 1\% accuracy loss, respectively. We manage to save more crossbar area compared with~\cite{wang2017group}, and reduce 96.65\% of the crossbar area for VGG-16 and 95.21\% for ResNet-18. The experimental results illustrate great potential for incorporating ADMM into structured weight pruning and quantization techniques on memristor-based DNN design, which will tremendously reduce the area and power consumption.

\subsection{Experimental Results on Structured Weight Pruning}
In our experiment, we compare our proposed framework with Group Scissor~\cite{wang2017group} as shown in Table \ref{table:results}.
% shows the comparison of our proposed pruning method with the Group Scissor~\cite{wang2017group}. 
Please note that we only prune CONV layers because they perform most of the FLOPs in the network calculation. On MNIST dataset, our original CNN model achieves 99.17\% accuracy, and 99.15\% accuracy with structured weight pruning. We also reduce our model size using extreme prune configuration, the results shows our method gets 98.33\% accuracy when we compress our model by 105.52$\times$.

On CIFAR-10 dataset, we construct different networks to test our method. Compared with the Group Scissor~\cite{wang2017group}, we not only achieve higher test accuracy using same compression ratio (2.35$\times$), but also manage to maintain same accuracy with a even higher compression ratio (2.93$\times$). For deeper network structures like VGG-16 and ResNet-18, we introduce such high regular sparsity into networks without accuracy degradation. Our framework reduces 13.98M and 10.46M weight volume for VGG-16 and ResNet-18 respectively.

% We also tested admm structured pruning method on ImageNet ILSVRC-2012 image classification. We noticed that the first convolutional layer of AlexNet is very small with only 35K weights compared with 2.3M in conv-2 to conv-5, and pruning conv-1 layer not only has no obvious effect on overall compression ratio but also hurts entire network performance. As a result, we skipped conv-1 when pruning are performed. Please note that even through we ignored pruning in one layer, we didn't skipped quantization in any layer which is for the sake of mapping weights on memristor. For the AlexNet, we achieved no accuracy degradation when we compressed model by 4.81$\times$, 0.7\% accuracy loss for 7.33$\times$ compression ratio and around 2\% accuracy loss for 15$\times$ compression ratio.

\subsection{Experimental Results on Weight Quantization for Memristor Crossbar Mapping}
\begin{figure} [t]
     \centering
%      \vspace{-1.2em}
     \includegraphics[width=1\columnwidth]{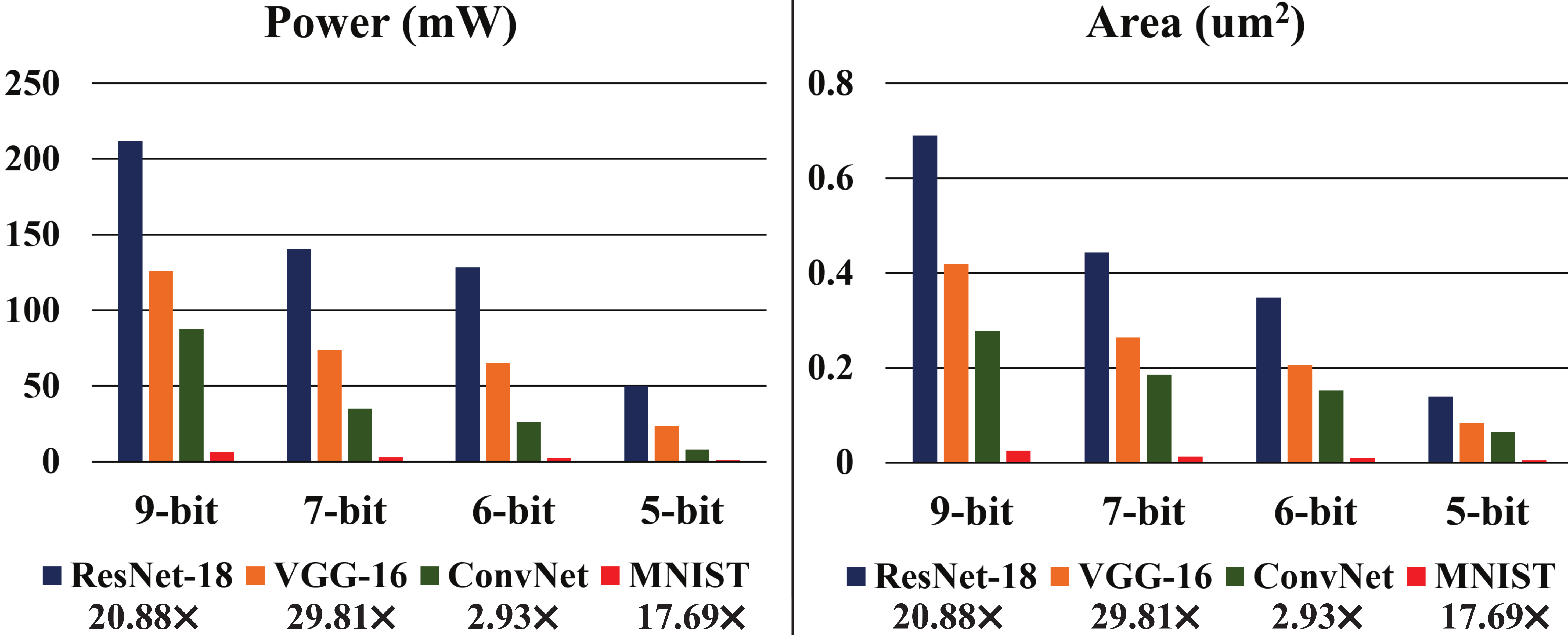}
     \caption{Total power and area reduction on compressed models using different quantization bits and networks}
     \label{fig:power}
\end{figure}

From the discussion in Section III-B.2, we can see that fewer bits can reduce the power as well as the memristor crossbar area. However, quantizing weights to some specific values will cause non-negligible accuracy degradation.
% when doesn't apply optimal algorithm for the quantization. 
In this paper, to mitigate accuracy degradation, we adopt ADMM to dynamically optimize well-leveled groups of weights which can be actually mapped on the memristors. By including memristor characteristics as discussed in Section III-B.2, our quantization process does not map weights to zero, and our state levels are zero-symmetric. Table \ref{table:results} shows different configurations for weight quantization and Figure \ref{fig:power} shows the power and area. Experimental results demonstrate that our framework maintains high weight prune ratio and fewer bits with promising test accuracy. According to 6-bit quantization results in Table \ref{table:results}, there is only 0.1\% accuracy degradation after quantizing LeNet-5 on 17.69$\times$ model, and only 0.2\% accuracy degradation after quantizing the 105.52$\times$ model. For a larger dataset such as CIFAR-10, a shallow ConvNet will introduce around 1.0\% accuracy degradation for our designed configuration (2.35$\times$) and 2.0\% accuracy degradation on a 5.88$\times$ compressed model, however as the network structure getting deeper, the accuracy drops around 0.1\% in a 9.31$\times$ compressed VGG-16 model and 0.5\% in a 11.75$\times$ compressed ResNet-18 model, and as the compression ratio gets larger, accuracy drops 0.8\% in a 29.81$\times$ compressed VGG-16 model and 0.6\% in a 20.88$\times$ compressed ResNet-18 model. 

As shown in Figure \ref{fig:power}, fewer bits represenation results in less power consumption and smaller area footprint, because the overhead of peripheral circuits such as ADCs and DACs will significantly decrease by lower the computing precision. There is a tremendous power and area reduction using the 5-bit quantization, since all memristor crossbars for higher bits representations are no longer needed. Beside the power and area reduction, fewer bits representation mitigates hardware imperfection of memristor including state drift and process variations. Compared to original DNN models, our 5-bit quantization models can achieve the largest power (area) reduction as 96.95\% (97.46\%), 98.38\% (98.28\%), 95.91\% (89.74\%) and 96.96\% (93.97\%) on ResNet-18, VGG-16, ConvNet and LeNet-5, respectively, among different bits representations.

% lower the requirement for memristor hardware which benefits the whole circuitry robustness.

\section{conclusion}
In this paper, we propose an unified and systematic memristor-based framework with both structured weight pruning and weight quantization by incorporating ADMM into DNNs training. 
%We are the first to consider the hardware constraints such as conductance range, and mismatch between weight value and real device into high structured sparsity model, and achieve high accuracy and low power and small area footprint at the same time. 
Three steps are mainly incorporated in our framework, which are \textit{memristor-based ADMM regularized  optimization}, \textit{masked mapping} and \textit{retraining}. We evaluate our proposed framework on different networks, and for each network, several pruning and quantizaiton scenarios are tested. On LeNet-5 and ConvNet, we can easily achieve better results comparing to Gourp Scissor. On VGG-16 and ResNet-18, after structured weight pruning and quantization, significant weight compression ratio, power and area reduction 5-bit weight representation can achieve significant power and area reduction network where only result in negligible accuracy loss, compared to the original DNN models.

\section*{Acknowledgment}
This work is funded by National Science Foundation CCF-1637559. We thank all anonymous reviewers
for their feedback.

% \begin{acks}
% \textcolor{red}{XXX}
% \end{acks}

\footnotesize
\bibliographystyle{IEEEtran}
% \bibliography{sample-bibliography}


%\renewcommand{\bibfont}{\footnotesize}
\renewcommand{\bibfont}{\small}

\end{document}